%% file: main.tex
\documentclass[sigconf,nonacm]{acmart}

\usepackage[caption=false]{subfig}
 \usepackage{float}
 \usepackage[normalem]{ulem}

\setcopyright{cc}
\setcctype{by-sa}

\title{Bring Buttons Back: Physical Interfaces for the Age of Automation}

\author{David Goedicke}
\orcid{0000-0002-4837-893X}
\affiliation{%
  \institution{Independent}
  \city{Würselen}
  \country{Germany}
}

\author{Donald Degraen}
\orcid{0000-0003-1029-931X}
\affiliation{%
  \institution{University of Canterbury}
  \city{Christchurch}
  \country{New Zealand}
}

\author{Tom Igoe}
\orcid{0009-0005-8379-0419}
\affiliation{%
  \institution{New York University (NYU)}
  \city{New York}
  \country{USA}
}

\author{David Sirkin}
\orcid{0000-0003-0134-6903}
\affiliation{%
  \institution{Stanford University}
  \city{Stanford}
  \country{USA}
}

\author{Wendy Ju}
\orcid{0000-0002-3119-611X}
\affiliation{%
  \institution{Cornell Tech}
  \city{New York}
  \country{United States}
 }

\author{Stefan Schneegass}
\orcid{0000-0002-0132-4934}
\affiliation{%
  \institution{University of Duisburg-Essen}
  \city{Essen}
  \country{Germany}
}

\begin{document}

\begin{abstract}

As AI starts to permeate everyday life, deployment focuses on embedding intelligence in the background, abstracting away controls, and leaving automated decision-making opaque, with few obvious opportunities for human intervention. Where interfaces remain, design has drifted toward abstract, screen-based control, replacing material interaction with menu navigation.
These shifts weaken the coupling between human action and system state, eroding operators’ ability to understand, anticipate, and intervene in automated systems.

We introduce Physically Stateful Interfaces (PSIs), a design concept that re-imagines traditional controls, such as buttons, switches, and knobs, as actuated, bidirectional interface elements. Three foundational behaviors, \textit{Self/Reset}, \textit{Resist/Hide}, and \textit{Assert/Unhide}, unify affordance, feedforward, and feedback into a single interaction point. Rather than static inputs whose impact and consequences must be read elsewhere on a screen, PSIs position physical controls as shared mediators between automated systems and human operators.

As automated systems become increasingly capable, their interfaces need to be deliberately designed to center human agency, supporting deliberate choice while keeping automation legible, contestable, and overrideable.

\end{abstract}
\maketitle

\input{Sections/01intro}

\input{Sections/02related}

\input{Sections/03ataxonomy}

\balance
\input{Sections/04discussion}
\section{Conclusion}
This work's contribution is not merely a prototyping exercise but a call to reorient (interface) design around people rather than technological allure. Early failures in automated systems highlight the consequences of failing to consider users and operators as integral parts of the systems they inhabit. As researchers, we must ensure that the systems we create embed a deliberate vision of the human–technology condition. 

Physically Stateful Interfaces are one exploration of how such a vision could take shape; in our case, bringing human-oriented physical controls into the age of automation to preserve human oversight, intuition, and agency. As technology becomes increasingly capable of acting independently, we must spend as much effort designing our relationship with it as we do designing the technology itself.

\subsection*{Acknowledgments}
\textbf{AI and Writing Tools.} ChatGPT and Grammarly were used to support proper use of language but the text was written by the authors. The authors reviewed and take responsibility for the final content.

\bibliography{references}
\bibliographystyle{acm}

\end{document}

%% file: Sections/01intro.tex
\section{Introduction}
Digitization and automation have revolutionized the objects and systems we encounter, imbuing them with capabilities to monitor, assist, and act autonomously. These pervasive computing devices, now supercharged with AI, infiltrate every corner of our lives.
Yet, in this pursuit of extreme functionality with AI and automation, system designs have often prioritized capability over interaction, overlooking how users engage with these systems. 
The simplicity and intuitiveness of physical controls have given way to abstract digital designs that prioritize manufacturing flexibility over usability. Touchscreens, capacitive sensors, and web interfaces have become dominant, often detaching control from the devices' physical context.
This abstraction, driven by cost optimization and technological allure, has come at the expense of intuitiveness and user agency, leaving us to question whether these changes have truly improved how we interact with the world around us.

In particular, AI-enabled systems need clear, tangible affordances, like buttons and switches, to keep their workings legible and under our control \cite{PowerButtonPlotnick}. As AI-driven technologies pervade our environment, research on interaction and automation must counter this trend and investigate responsible solutions that maintain human oversight and agency.

\subsection{Eroding Human Agency}
These issues persist across domains, from nuclear control rooms to naval navigation to modern automobiles; abstracted and detached interface design has long shown its potential to introduce complications into critical automated systems. Long before web interfaces and touchscreens, technology-driven design de-prioritized human oversight. The 1979 \textit{Three Mile Island} (TMI) nuclear accident exemplifies this.

At TMI, operators misinterpreted key indicators because the interface obscured the reactor's actual state. Disorienting warning lights and alarms failed to convey critical pressure and coolant levels, while a single-line printer delivered status updates too slowly for real-time decision-making~\cite{cringelyThreeMile,threeMileIslandWiki}.


While automation at TMI was introduced for safety and efficiency, it followed a \textit{design philosophy that sidelined human control}, assuming failures could be fully anticipated. As Elish describes in \textit{Moral Crumple Zones}~\cite{elish2019moral}, automation often shifts responsibility to human operators only at the moment of failure—without providing the tools for effective intervention.

Ensuring human operators retain meaningful control must be a priority, especially as AI and automation expand. Although design methods have evolved, the core issue remains: interaction design is often an afterthought in system design. This challenge is particularly pressing as touchscreens and web interfaces continue to replace physical controls in safety-critical environments.


The \textit{USS John S. McCain} collision demonstrates the risks of touchscreen-based interface systems in critical contexts. Steering and throttle controls, buried within digital menus, obscured key parameters from operators, leading to unintended actions and delayed responses. Reliance on intangible controls, compounded by inadequate training, led to miscommunication and loss of situational awareness, highlighting the advantages of direct physical control for immediate, reliable interaction~\cite{ussMcCainCrash}.


Similar risks are found in everyday interactions with automation. The reliance on touch interfaces in cars poses significant risks. In 2024, \textit{Volkswagen ID.4} owners reported unintended acceleration due to touch-sensitive steering wheel controls~\cite{vwID4Touch, Wired2025_RejoiceButtons}. Recognizing these risks, Euro NCAP announced that vehicles must include physical controls for selected essential functions to qualify for its highest safety rating~\cite{euroNCAPTouch}. 

These are not merely technological issues but fundamental design failures, where cost pressures often eclipse usability and intuition. 

Yet, designs that prioritize usability and intuitiveness often emerge from simple yet deliberate refinements in materiality rather than added complexity. Designs like the OXO kitchen tools illustrate the power of simple, thoughtful material improvements. Ergonomic handles transformed everyday utensils into more inclusive and intuitive tools without the need for digital complexity. Yet, as a field, we often neglect these lessons, favoring technological novelty and user-agency-reducing automation while overlooking the benefits of legacy physical interfaces.

Outside academia, the maker community has explored similar ideas, introducing modern
functionality into legacy devices through mechanical retrofits. Examples can
be found on maker platforms such as \href{https://www.instructables.com/Automatic-Light-Switch-4/}{Instructables}
or from community creators such as \href{https://www.instagram.com/cartyski/}{Carty Ski}.

While digital abstraction dominates modern interfaces, legacy physical controls remain essential where clarity and intuitive operation, immediate access, and intervention matter. Rather than discarding these elements, research must examine how successful interactions emerge, what makes them effective, and how to rediscover them for future automated and AI-supported designs.

%% file: Sections/02related.tex
\section{Insights from past Research}
The failures of \textit{Three Mile Island}, the \textit{USS McCain}, and the \textit{Volkswagen ID.4} illustrate a fundamental breakdown in interaction design. These failures did not result solely from technological shortcomings but from critical gaps in how users could execute actions and evaluate system states. Don Norman~\cite{norman1988design} describes these challenges as the \textit{Gulf of Evaluation}, where users cannot easily perceive or interpret the system’s state, and the \textit{Gulf of Execution}, where users struggle to determine how to act on a system. 

At \textit{Three Mile Island}, fragmented alarms and the slow, single-line printer delayed operators' ability to assess system status, exacerbating the \textit{Gulf of Evaluation}. The \textit{USS McCain} disaster resulted from an overcomplicated touchscreen interface that buried essential controls, exacerbating the \textit{Gulf of Execution}. The \textit{Volkswagen ID.4's} capacitive controls remove tactile feedback, forcing users to rely on visual confirmation and increasing both gulfs.

Human-computer interaction (HCI) research has developed frameworks to bridge these gulfs. In the following sections, we examine the concepts used in these frameworks, specifically affordances, feedforward, and feedback. 

\subsection{Affordance}
Affordances, first introduced by James Gibson in ecological psychology~\cite{gibson2014ecological}, describe the action possibilities an environment offers an organism. Don Norman~\cite{norman1988design} later extended this concept to design, emphasizing that interfaces should visibly communicate how they can be interacted with (i.e., showing affordances). Users evaluate affordances early in interaction, and designs that account for this can support intuitive onboarding and seamless engagement.

Gaver introduced the concept of \textit{perceptible}, \textit{hidden}, and \textit{false} affordances, emphasizing that not all affordances are directly communicated through an object’s design; some must be inferred or discovered through use~\cite{gaver1991technologyaffordances}. Hartson later expanded on this framework by defining \textit{cognitive}, \textit{physical}, and \textit{functional} affordances, which highlight the role of feedback and the tight coupling between input and system state~\cite{hartson2003affordances}. Affordances make interfaces discoverable and intuitive, leveraging our physical embodiment to guide user action.


\subsection{Feedforward}
Feedforward describes an interface's ability to signal what will happen after an interaction. As Vermeulen et al. discuss, feedforward can take many forms, including physical constraints, visual cues, and dynamic responses that guide user action~\cite{revealingFeedforward}.

\subsection{Feedback}
Feedback signals a system’s response to user actions, enabling them to verify input success and interpret the system state. Clear, immediate feedback builds trust and supports intuitive interactions, extending beyond single actions to convey overall system status.

Research on intelligible systems~\cite{explainableSystems} and home automation~\cite{homeAutomationResearch} highlights the importance of designing interfaces that externalize internal system states, especially when hidden control elements can change them.

Modern interfaces fragment direct feedback and feedforward communication. Touchscreens, for example, separate input and feedback into different locations, increasing cognitive load and slowing reaction times~\cite{colley2019touchscreens}.

\subsection{Integrating Interaction Principles}
While affordances, feedforward, and feedback are often examined separately, effective design frameworks integrate them to create seamless and intuitive interactions. Winograd and Flores~\cite{winograd1986understanding} exemplify these ideals, building on Heidegger’s concept of \textit{ready-to-hand} and arguing that well-designed technologies should fade into the background, allowing users to focus entirely on their tasks and act fluidly through the tools they use.

Building on this goal, Dourish and Ju explore how to synthesize these concepts into coherent interaction designs. Dourish~\cite{Dourish2001Action} introduces \textit{embodied interaction}, arguing that interfaces should align with human practices rather than imposing abstract control schemes. Ju~\cite{ju2015design} expands on these ideas with \textit{implicit interaction}, where system and user behavior cues subtly guide the interaction, avoiding explicit attention and interaction. When implemented effectively, these approaches reduce cognitive load and enhance usability, particularly in automated environments.

\subsection{Tangible Interface Research Prototypes}
Tangible User Interfaces (TUIs) have often been critiqued for their perceived lack of flexibility compared to graphical user interfaces (GUIs) \cite{10.1145/3544549.3582744}. This critique underestimates the potential of well-designed physical interfaces to adapt dynamically to user needs while overlooking how physical constraints can encourage deliberate choices about system behavior. Research in reconfigurable and shape-changing technologies demonstrates that tangibility and adaptability are not mutually exclusive. Understanding how physicality contributes to interaction design requires examining research at multiple levels, from designing individual interface elements to broader interaction frameworks and adaptable physical controls.

Interaction primitives (e.g., buttons and switches) often inherently embody key design qualities mentioned above. Oulasvirta et al. \cite{oulasvirta2018neuromechanics} examined the design space of these primitives, demonstrating how tactile feedback in button design enhances precision and ergonomics. Van Deurzen et al. \cite{vandeurzen2024substitutebuttons} explored how a limited set of physical buttons can effectively substitute for a wide range of tactile interactions, identifying six substitute buttons that emulate various button designs. These works highlight the vast design space within the most fundamental interaction elements.

By combining these primitives, we can create more flexible designs while preserving their inherent qualities. Notable examples include dynamically adapting form and function to better support specific applications. Tiab and Hornbæk~\cite{tiab2016shapechangingbuttons} demonstrated how shape-changing buttons enhance affordances and feedback, enabling more nuanced user interactions. Similarly, Kim et al.'s \textit{KnobSlider} and \textit{ExpanDial} devices illustrate how physical controls can transition between interaction modes—such as a slider becoming a knob—while maintaining tactile clarity and usability~\cite{kim2018knobslider, kim2019expandial}.

Beyond individual components, tangible interaction research explores how materiality and computational adaptability merge to create more holistic interaction paradigms. Ishii et al.'s concept of \textit{Radical Atoms}~\cite{ishii2012radical} envisioned computationally controlled physical materials that can dynamically transform their shape to represent digital information. The project inFORM~\cite{follmer2013inform} extended this idea by enabling a physical device to provide both input and output. An established implementation of related principles appears in \textit{motorized faders on audio mixing consoles}, where the automated movement of physical sliders provides \textit{real-time, tactile feedback} directly coupled to the system state. Similarly, \textit{adaptive tactile displays} in assistive technologies help users interact with data meaningfully~\cite{Jafari2016-jd}.

These innovations show that, when implemented thoughtfully, TUIs can generate innovative interaction concepts. TUIs can be flexible, but they can also leverage physicality to reduce cognitive load, provide intuitive mappings, and maintain direct control while forcing designers to make confident choices about a system's use and features.

To bridge Norman's gulfs of evaluation and execution, interface designs should collapse traditionally separate characteristics into unified designs that externalize system states and guide users seamlessly. The following sections explore how legacy interfaces can be re-imagined to play a central role in the future.

%% file: Sections/03ataxonomy.tex
\section{Physically Stateful Interfaces}
In this work, we introduce the term \textit{Physically Stateful Interfaces} (PSI) to describe physical controls that function as unified, stateful elements, combining input and output in a single interaction point. By leveraging the inherent advantages of physicality—tactile feedback, intuitive affordances, and material clarity — PSIs offer a structured approach to reintroducing physical controls in the age of automation.

\textbf{The PSI concept is guided by a central question:}

\textit{What functionality could physical interactive elements (e.g., buttons, switches, and knobs) incorporate to remain viable for controlling automated systems while preserving their inherent advantages?}

From this question, we identify four design considerations:
\begin{enumerate}
    \item \textbf{Affordance:} What actions does the physical control make available and apparent?
    \item \textbf{State Legibility:} How does the physical state of the interface communicate the state of the system?
    \item \textbf{Bounded Control:} How does the interface communicate what users can, cannot, or should do, including opportunities for deliberate intervention and override?
    \item \textbf{Shared Control:} How can both the automated system and the human act through the same physical interface while preserving a coherent state?
\end{enumerate}


Building on these considerations, we propose a taxonomy of three core behaviors that physical interfaces can adopt to enhance control and adaptability:

    \begin{figure}[ht]
    \centering
     \caption{Visual sequence of a rocker switch
changes state through an automated system.}\label{fig:SelfFigure}%
    \includegraphics[width=0.9\linewidth]{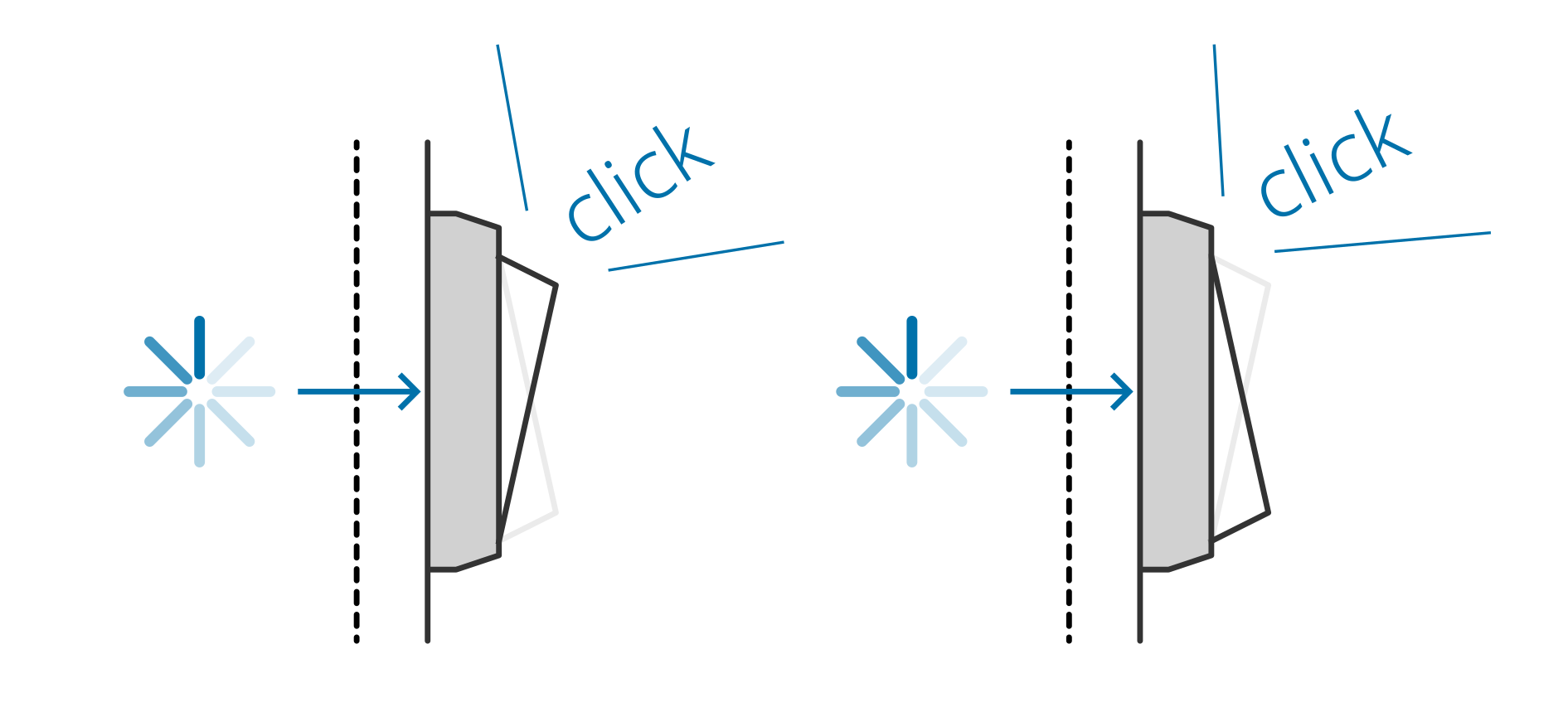}
    \Description{Pictogram representing PSI-Self. A light switch whose physical position is changed automatically
by the system when the light is switched remotely.}

        \end{figure}
\begin{enumerate}
    \item \textbf{Self \& Reset}: These interfaces autonomously change state or revert to a default state, aligning their physical configuration with the system's status while still allowing user control.
    \item \textbf{Resist \& Hide}: Interfaces may impose physical constraints or become inaccessible under certain conditions; these limitations might be used to ensure safe operation and might be user-overridable.
    \item \textbf{Assert \& Un-Hide}: Interfaces that guide users by asserting or un-hiding specific controls to assist in decision-making by indicating subsequent or urgent actions.
\end{enumerate}

Examples such as flying faders on digital mixers or resettable fuses demonstrate existing behaviors. PSIs offer a path to extend these ideas by reimagining how physical elements can integrate into automated systems for stateful, dynamic control.

The following section explores the design space in more detail through short examples of each type.

\input{Sections/03bTaxonomyTable}

%% file: Sections/03bTaxonomyTable.tex
  \begin{figure}[htbp]
  \centering
  \caption{Visual sequence of a latching push-button that digitally de-latches.}\label{fig:ResetFigure}%
    \includegraphics[width=0.90\linewidth]{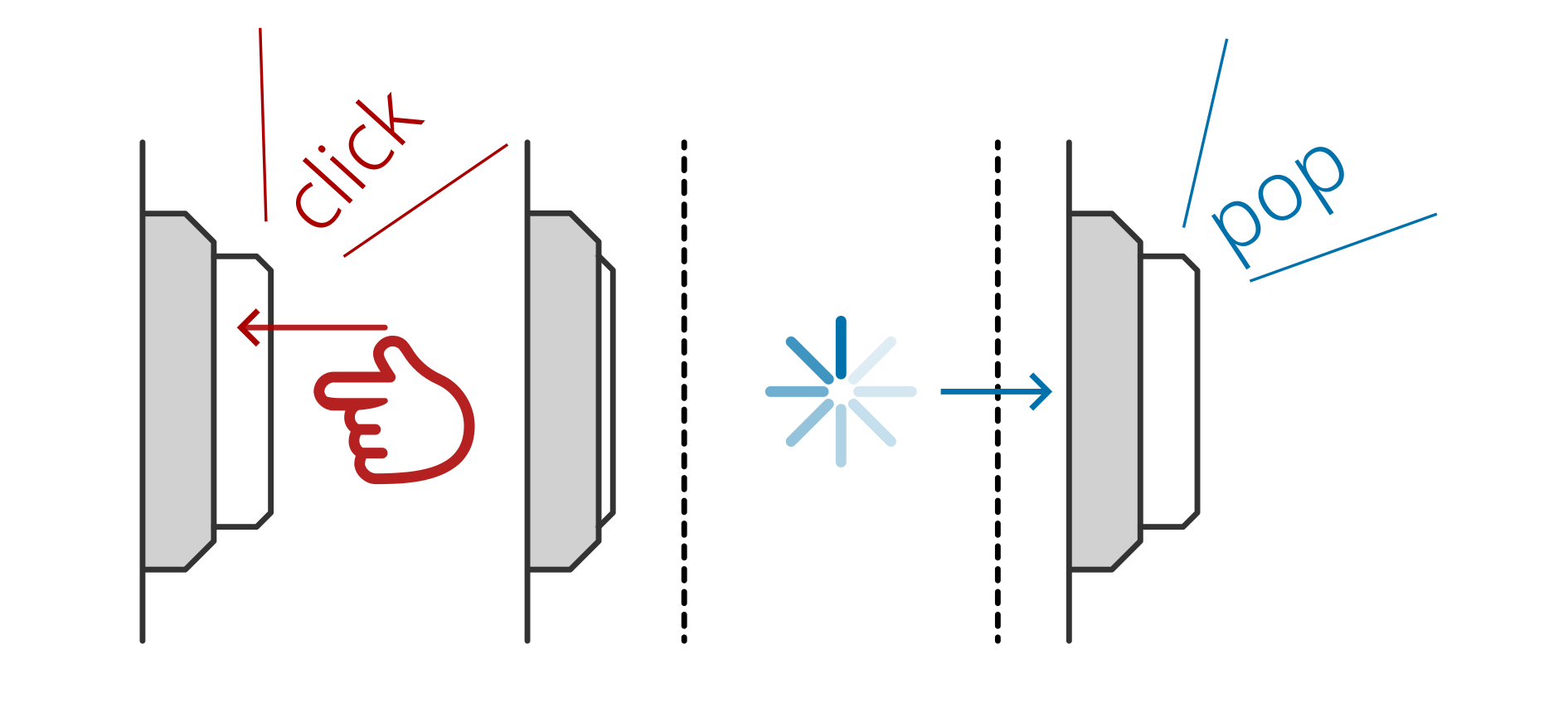}
    \Description{Pictogram representing the PSI Reset. A toggle button that resets automatically when a triggered task is complete e.g. after establishing a network connection.}
  \end{figure}

\subsubsection*{Self \& Reset}
The self-switching PSI lets the system control an interface element's state, allowing it to toggle or reset based on subsystem conditions. In some cases, the system may only need to unlatch a maintained switch upon task completion or when a condition is met; in others, it may completely alter the switch's state. See: \autoref{fig:SelfFigure} \& \autoref{fig:ResetFigure}.

In this way, a PSI's state can be tightly coupled with the controlled element, serving as both a control input and a real-time indicator of system state.

\textbf{Self-switching Example: Home Automation}
In home automation, the self-switching PSI can control a light switch, with automated rules managing its state. The system can toggle the switch on or off based on predefined conditions, while user override remains as natural as using a traditional switch. See \autoref {fig:SelfFigure}.

\textbf{Reset Example: Establishing a network connection}
The reset PSI can communicate the state of a longer task, such as establishing a remote connection. The device attempts to connect while the button remains latched and automatically unlatches once it connects. See Figure~\ref {fig:ResetFigure}.

\subsubsection*{Resist \& Hide}
When a system detects an unsafe or inadvisable action, a PSI could resist its operation or hide itself.

\begin{figure}[htbp]
\centering
    \caption{Visual sequence of a push button that resists activation until the system changes the state, removing the resistance and allowing activation.}\label{fig:ResistFigure}
    \includegraphics[width=0.90 \linewidth]{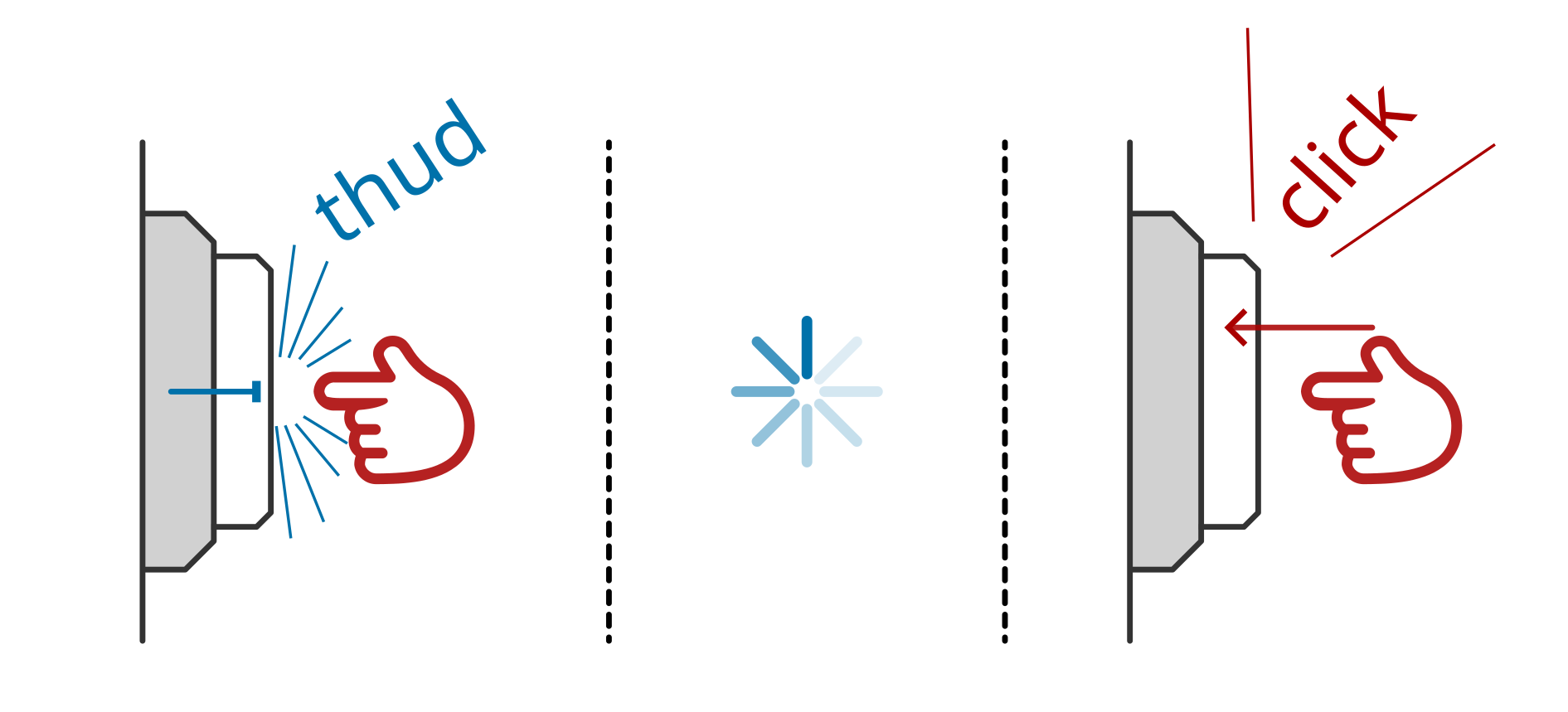}
    \Description{Pictogram representing the PSI Resist. A physical control that first resists to be pressed and later changes its resistance such that pressing becomes possible.}
  \end{figure}

\begin{figure}[htbp]
\centering
  \caption{Visual sequence of a pushbutton that is hidden by the system, when its operation would be unsafe.}\label{fig:HideFigure}
  \includegraphics[width=0.90 \linewidth]{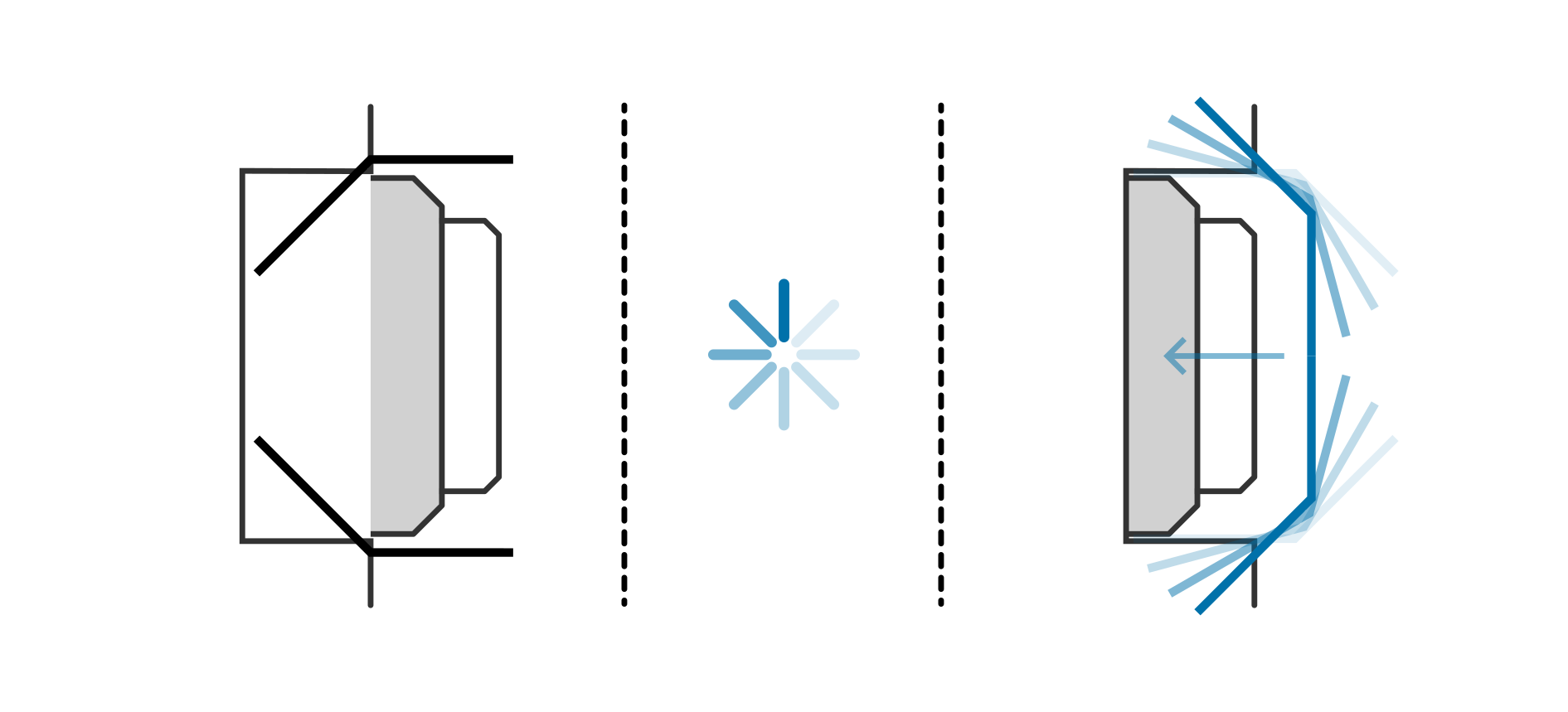}
  \Description{Pictogram representing the PSI Hide. A physical control that becomes in accessible, based on a systems decision.}
\end{figure}

This resistance may fully prevent operation or allow users to override it with additional force. In either case, the interface resists actions until conditions are deemed safe, aligning them with operational safety requirements. See~\autoref{fig:ResistFigure}~\&~\autoref{fig:HideFigure}.

\textbf{Example: Power Generator Synchronization}
A resistive button could control a power generator's connection to the power grid. As phase and frequency become synchronized, the button becomes progressively easier to press. This gives operators clear feedback on when to connect the generator while preserving user agency and control. See~\autoref{fig:ResistFigure}.

\textbf{Example: Convertible Car Roof}
A button to lower the roof of a convertible could retract and hide when the car is shifted into gear or begins moving, preventing unsafe operation. Besides enhancing safety, this design communicates the relationship between user inputs (driving the car) and safety constraints. See~\autoref{fig:HideFigure}.

\subsubsection*{Assert \& Un-Hide}
When a system identifies an optimal next action or emergency response, it can use interface elements to implicitly suggest that action to users.

\begin{figure}[htbp]
    \centering
   
    \caption{Visual sequence of a push-button being raised from its flush position to suggest an action to the user.}\label{fig:AssertFigure}{\includegraphics[width=0.90 \linewidth]{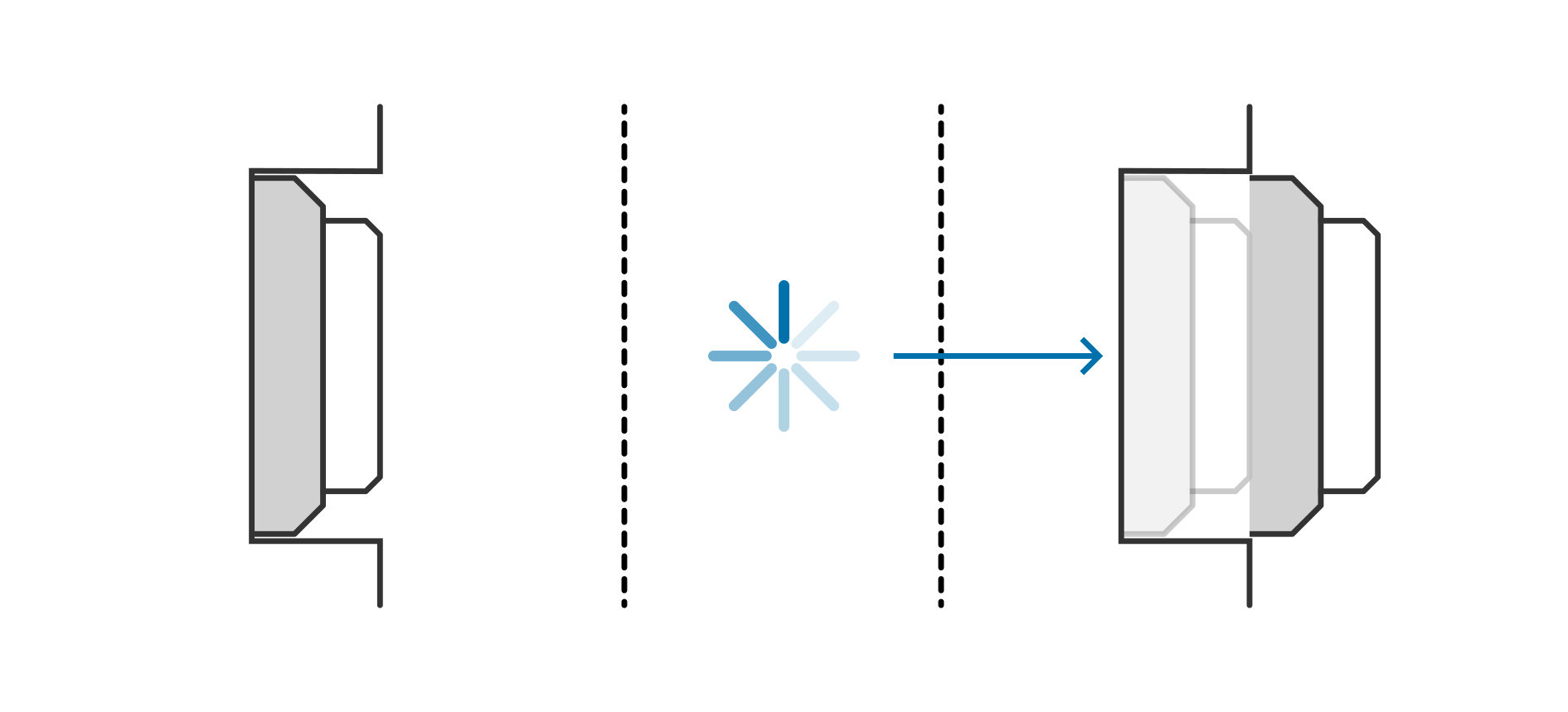}}
\Description{Pictogram representing the PSI Assert. A physical control that is moved to become visually and tactually more prominent based on a system decision.}
\end{figure}

\begin{figure}[htbp]
 \centering
\caption{Visual sequence of a system opening a cover over an emergency stop button, making the control more prominent and signaling uncertainty to the user.}\label{fig:UnHideFigure}{\includegraphics[width=0.90 \linewidth]{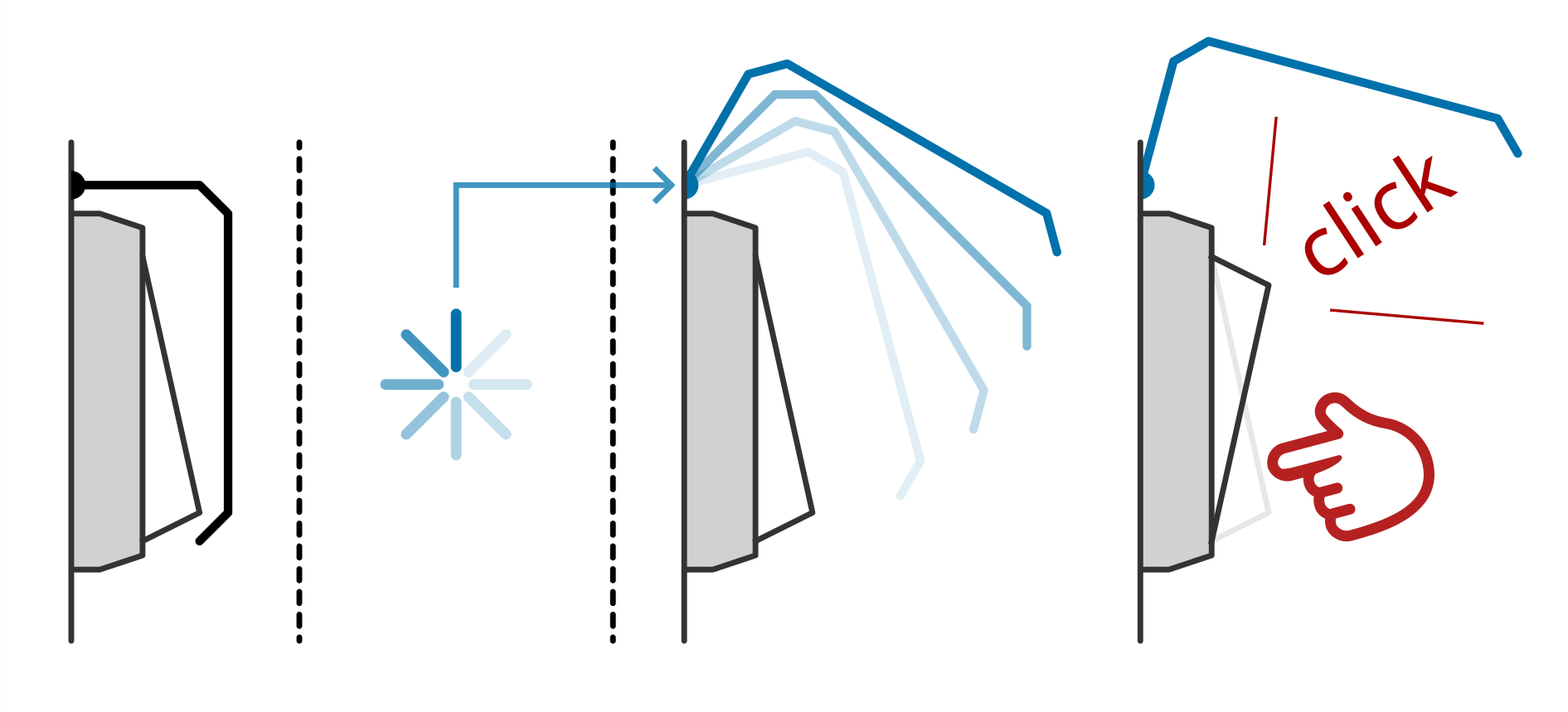}}
\Description{Pictogram representing the PSI Un-Hide. An emergency control that becomes physically accessible when the system determines that operator intervention is required.}
    \end{figure}

It could raise a button or remove a protective cover to make the interface noticeable. This mechanism maintains operator control while letting the system suggest subsequent actions to support decision-making. See Figure~\autoref{fig:AssertFigure}~\&~\autoref{fig:UnHideFigure}.

\textbf{Example: Manufacturing Application}
In manufacturing, where workflows are sequential, the system can suggest the next step by asserting the corresponding control. For example, it might raise a button to indicate that a material is ready for the subsequent stage.
This aligns system predictions with user actions while preserving operator control. See Figure~\ref {fig:AssertFigure}.

\textbf{Example: Industrial Application}
In a chemical processing plant, systems often operate near critical thresholds such as temperature or pressure limits.
If conditions become unstable but remain within safe parameters, the system could make the emergency stop button visually and physically prominent. This lets operators recognize the system's uncertainty and take manual action if needed, maintaining oversight by those responsible for control. See Figure~\ref {fig:UnHideFigure}.

%% file: Sections/04discussion.tex
\section{Closing Thoughts}
Precise control of volatile systems is crucial, especially as automation and AI expand across an increasing range of domains. PSI integrates human-legible properties such as affordances, feedforward, feedback, and statefulness into a single interaction point, reinforcing operators' agency to intervene.

Furthermore, the physicality of such interfaces imposes deliberate constraints that require designers to make decisions about control and responsibility before deploying a system. With PSIs, designers must determine which actions are available to users, which actions automation may perform, and when either can override the other.

Existing projects have already explored ideas closely related to PSI. For example, RetroFab~\cite{ramakers2016retrofab} shows the potential of integrating physical and digital systems by introducing stateful, tactile interfaces to legacy devices.

\textbf{Video Games} \\
Examples also appear in video game design. Games often use familiar physical controls, such as buttons, switches, and levers, to show how players can interact with the virtual world. Because these controls are virtual, their physical state can also change dynamically to communicate system state or suggest subsequent actions. This makes games a compelling example of how familiar physical interfaces can guide action and form the basis for intuitive interface design.

Since we first developed this concept in 2022, games such as \href{https://store.steampowered.com/app/2956040/PVKK_Planetenverteidigungskanonenkommandant/}{PVKK} and \href{https://store.steampowered.com/app/2950790/IRON_NEST_Heavy_Turret_Simulator/}{IRON NEST} have further explored this interaction paradigm, using stateful controls, feedforward, and feedback to guide players through complex operational procedures with minimal explicit instruction.


\textbf{Electronic Instruments} \\
The synthesizer community offers another compelling example.
Designers and musicians in this field have long embraced the "one button, one function" (OBOF) principle, where each control serves a specific function. This design philosophy prioritizes clarity and precision, letting musicians easily build intuition with an instrument without navigating complex menus. 

\begin{figure}[H]
    \centering
    \includegraphics[
    trim=0 200 0 200,
    clip,
    width=0.90\linewidth]{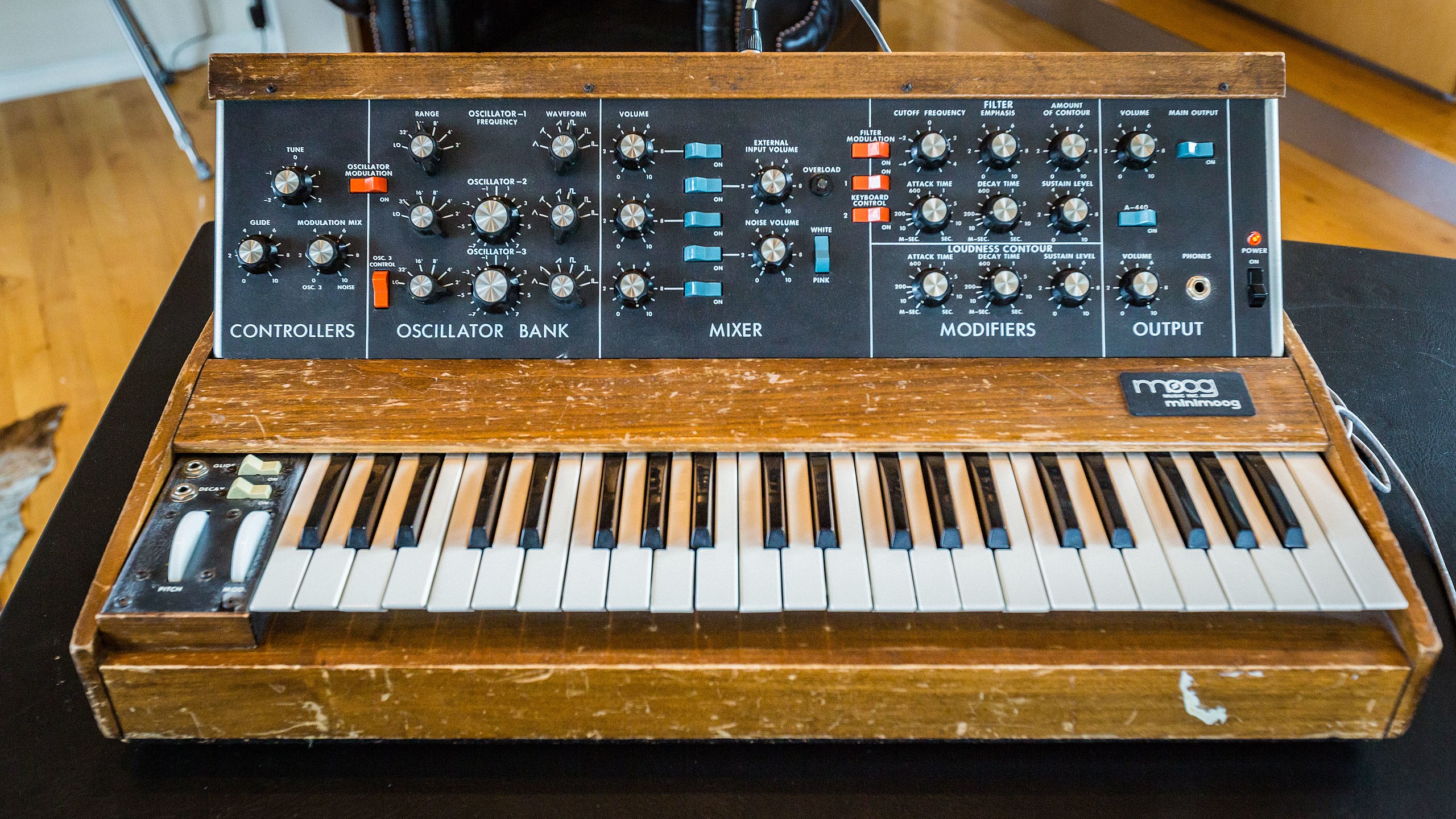}
    \caption{An iconic Minimoog synthesizer renowned for its array of dedicated physical controls. \textit{\small Image Credit: Ville Hyvönen/Wikimedia Commons, CC BY-SA 2.0}}
    \label{fig:ModelD}
    \Description{A photograph of an original Minimoog synthesizer, showing the keyboard below and dedicated knobs and switches above for controlling the timbre of the sound, illustrating the One Button, One Function principle.} 
\end{figure}

PSI extends the OBOF principle into automated systems, maintaining core tenets while introducing dynamic control states. Clear states and a limited set of important physical controls reduce cognitive load, fostering intuitive, instinctive use.
See Figure~\ref{fig:ModelD} for an example of an OBOF instrument.

\subsection{Future Work}
\begin{figure}[htbp]
    \centering
    \caption{A push-button prototype that can digitally de-latch and resist a user's input.}\label{fig:DeLatchPrototype}
    \includegraphics[
    trim=0 200 0 10,
    clip,
    width=0.45\linewidth]{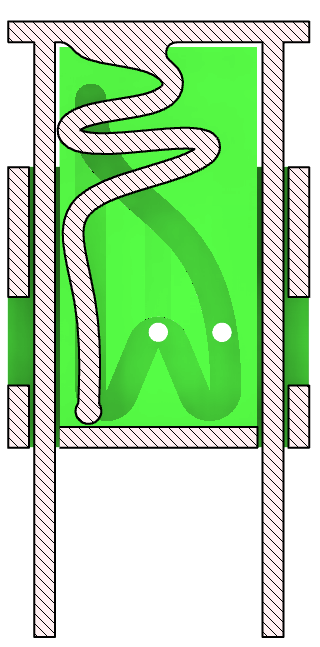}
    \Description{A cross-sectional view of a mechanical latching mechanism, similar to that found in a ballpoint pen, with additional cutouts that enable the Reset and Resist interaction modalities.}
    
\end{figure}

PSI currently exists primarily as a design concept supported by early prototypes (see ~\autoref{fig:DeLatchPrototype}). Our initial explorations show that familiar controls can be retrofitted to include the behaviors described above. However, the broader design space remains largely unexplored, and replication remains a challenge.

Materiality is an essential property of these interface elements, which implies that before we can establish their efficacy, we should refine the electromechanical design to be robust and replicable. To this end, we invite anyone to contribute ideas, concepts, or devices on \href{https://github.com/SpeculativePrototyping/Physically-Stateful-Interfaces}{and on GitHub.}

